
\magnification=1200
\vbadness=10000
\hfuzz=10pt \overfullrule=0pt
\baselineskip=12pt
\parindent 20pt \parskip 6pt
\def\O{{\cal O}}
\def\E{{\cal E}}

\def\eh{{\hat{\bf e}}}

\def\ee{\eta_\E}
\def\Eb{\bar\E}
\def\dh{{\hat{\bf d}}}
\def\x{{\bf x}}

\def\p{{\bf p}}
\def\0{{(0)}}
\def\1{{(1)}}
\def\2{{(2)}}
\def\G{\Gamma}
\def\dT{\widehat{\delta T}}

\hfill {} \vskip .3in

\centerline{{\bf Higher-Order Gravitational Perturbations}}
\vskip .2cm
\centerline{{\bf of the Cosmic Microwave Background}
\footnote{$^*$}{This work was supported in part by
the National Science Foundation under grants AST/90-05038 and
PHY/9200687, by NASA
contract NAS8-39073, and by the U.S. Department
of Energy (D.O.E.) under cooperative agreement DE-FC02-94ER40818.}}
\vskip .3in

\centerline {Ted Pyne$^{(1)}$  and Sean M.~Carroll$^{(2)}$}
\vskip .3cm
\centerline{\it $^{(1)}$Harvard-Smithsonian Center for Astrophysics}
\centerline{\it Cambridge, Massachusetts\quad 02138}
\centerline{{\it email: pyne@cfa160.harvard.edu}}
\vskip .3cm
\centerline{\it $^{(2)}$Center for Theoretical Physics, Laboratory for
Nuclear Science}
\centerline{\it and Department of Physics}
\centerline{\it Massachusetts Institute of Technology}
\centerline{\it Cambridge, Massachusetts\quad 02139}
\centerline{{\it email: carroll@ctp.mit.edu}}

\vskip .3in

\centerline{\bf Abstract}
\vskip .1in

We study the behavior of light rays in perturbed Robertson-Walker
cosmologies, calculating the redshift between an observer and the
surface of last scattering to second order in the metric perturbation.
At first order we recover the classic results of Sachs and Wolfe, and
at second order we delineate the various new effects which appear;
there is no {\it a priori} guarantee that these effects are
significantly smaller than those at first order, since there are large
length scales in the problem which could lead to sizable prefactors.
We find that second order terms of potential observational interest
may be interpreted as transverse and longitudinal lensing by foreground
density perturbations, and a correction to the integrated Sachs-Wolfe
effect.

\vfill

\centerline{CTP~\# 2455 \hfill October 1995 (minor revisions December 1995)}

\centerline{astro-ph/9510041 \hfill}
\vskip .1in

\eject
\baselineskip=16pt

\noindent{\bf I. Introduction}

\nobreak
In the last several years, observations of temperature anisotropies
in the cosmic microwave background (CMB) [1] have spurred
increasingly sophisticated investigation of the
anisotropy predicted by theoretical models [2-5].  Important
contributions to the anisotropy come from gravitational
perturbations, temperature and pressure fluctuations at the
surface of last scattering, and ionization effects in the later
universe.

The earliest of these effects to be studied, and the most important
on large scales, are those due to gravitational perturbations.
These were systematically investigated by
Sachs and Wolfe [6], who derived the basic
formulae relating perturbations in the metric to anisotropy
in the temperature of the CMB.  Their results revealed two basic
sources of anisotropy: potential fluctuations at the surface of
last scattering, and time variation of the potential along the
path of the photon.  Later investigations focused
on individual effects in specific models [7-15].

Even though perturbations in the energy density $\delta\rho/\rho$
grow to be greater than unity on sufficiently small scales, the
resulting metric perturbations may almost always be taken to be small
[16].  It therefore makes sense to calculate the behavior
of photons to first order in this pertubation, as Sachs and
Wolfe did.  Nevertheless, there is no way of knowing ahead of time
that second-order terms in an expansion in the metric perturbation will
be negligible compared to the first order terms, since there is ample
opportunity for effects to accumulate as photons travel to the observer
from the surface of last scattering; in other words, the coefficients
of the second-order terms may be numerically large.  (As an example
of a related effect, the time delay formula in standard gravitational
lens systems contains important contributions from both the first order
Shapiro and the second order geometric effects.)
It is therefore worthwhile investigating
the redshift induced by effects which are formally second order
in the metric perturbation to see if they may nevertheless be
observationally important.  In this paper we calculate these
second-order effects and interpret the results in terms of
specific physical processes.

It is necessary to be careful about what we mean by ``second
order'' in the context of gravitational perturbation theory.
We imagine that we are given a metric throughout spacetime of
the form
  $$g_{\mu\nu} = g^{(0)}_{\mu\nu} + h_{\mu\nu}\ ,\eqno(1.1)$$
where $g^{(0)}_{\mu\nu}$ describes a background Robertson-Walker
spacetime and $h_{\mu\nu}$ is a small perturbation.  We will
not be computing this perturbation to second order
in pertubations of the energy-momentum tensor,
but simply calculating photon trajectories to second order
in $h_{\mu\nu}$ and its derivatives.
Therefore, if $h_{\mu\nu}$ is computed from standard first-order
metric perturbation theory and substituted into our expressions,
the results will not represent a complete calculation of
effects which are second order in the matter perturbations.
(In Sec.~IV we will examine explicitly the case of
first-order scalar perturbations, but it is straightforward
to generalize the results.)
Nevertheless the expressions we obtain will constitute
a subset of all the possible contributions,
and if any of them turn out to be comparable in magnitude to
terms which are formally first order, it is appropriate to
take them into account. Moreover, the substitution
$h_{\mu\nu}\mapsto g^{(1)}_{\mu\nu}+g^{(2)}_{\mu\nu}+\dots $
into our formulae below would immediately yield an expansion for
the full second order anisotropy.

It is also important to note that we will only be dealing with
gravitational perturbations.  We will imagine that there is
a hypersurface of last scattering fixed at some definite time,
on which there can exist intrinsic perturbations which may
be calculated independently; we then
compute the additional perturbations due to the metric
fluctuations along the geodesics followed by the photons.
Non-gravitational second-order perturbations were treated by
Vishniac [17], Dodelson and Jubas [18], and Hu, Scott and
Silk [19]. The latter authors also examined higher order
gravitational effects by expanding the Boltzmann equation
to second order, but did not construct explicit solutions.
As a final caveat, we treat the order-by-order expansion in
powers of the metric perturbation and its derivatives in a
formal sense; thus, a phenomenon such as the integrated
Sachs-Wolfe (or Rees-Sciama) effect
we consider to be first order (since it involves terms linear in
derivatives of $h_{\mu\nu}$), even though it is sometimes thought
of as second order since it can be numerically small (and vanishes
to first order in some specific models).

Our calculation proceeds as follows.  In Sec.~II we set up
the problem and express the redshift experienced by a photon
in terms of its corresponding background path $x^{(0)\mu}(\lambda)$
and its first and second order perturbations, $x^{(1)\mu}(\lambda)$
and $x^{(2)\mu}(\lambda)$.  In Sec.~III we discuss a general
formalism for constructing these pertubations in terms of the
metric variables; this is an extension of the methods of
Pyne and Birkinshaw [20]
to arbitrary order.  In Sec.~IV we specialize to the case of
scalar pertubations, and examine the resulting formula for the
temperature anisotropy.  Although a quantitative understanding
of the magnitude of each term would require detailed knowledge
of the evolution of the metric perturbations (which we do not
attempt in this paper), it is possible to discuss informally which
contributions might be observable in realistic models of
structure formation.

\vskip .1in

\noindent{\bf II. Perturbation Expansion}

\nobreak
We are interested in the pattern of temperature fluctuations
$\Delta T/T$ on the sky as seen by an observer in a perturbed
Robertson-Walker spacetime.  We write our background metric in
conformal coordinates $x^\mu=(\eta,x,y,z)$ as
  $$\eqalign{d\bar{s}^{2(0)} &= \bar{g}^{(0)}_{\mu\nu}
  dx^\mu dx^\nu \cr
  &= a^2(\eta)[-d\eta^2 +\gamma^{-2}
  (dx^2 + dy^2 + dz^2)] \ .\cr }\eqno(2.1)$$
Here $\gamma=1+\kappa r^2/4$, where $\kappa$ is the spatial
curvature parameter ($+1$, $-1$ and $0$ for positively curved,
negatively curved and flat cases, respectively),
$a(\eta)$ is the scale factor, and
$r^2=x^2+y^2+z^2$.  In this section we
consider an arbitrary metric perturbation $\bar{h}_{\mu\nu}$.
It will be convenient to separate out the
dependence on the scale factor by working in the conformal
background metric $g^{(0)}_{\mu\nu} = a^{-2}\bar{g}^{(0)}_{\mu\nu}$,
with the conformally-transformed perturbation
$h_{\mu\nu} = a^{-2}\bar{h}_{\mu\nu}$, so that the actual,
physical spacetime metric is given by ${\bar g}_{\mu\nu}
={\bar g}^{(0)}_{\mu\nu}+{\bar h}_{\mu\nu}$.
The wavevector $\bar{k}^\mu$
of a light ray in the physical metric is related to the wavevector
${k}^\mu$ in the conformally transformed metric by  $k^\mu =
a^2\bar{k}^\mu$.  (Our conventions are those of Ref.~[20].)

Within such a spacetime we consider a photon path $x^\mu(\lambda)$,
where $\lambda$ is an affine parameter.  (See Fig.~1.)
This path connects an observer at a point $\O$ with
coordinates $x^\mu_\O=(\eta_{\O},0,0,0)$
to the hypersurface of emission, which we define to
be the spacelike hypersurface of constant conformal time $\eta=\ee$.
The ``surface'' of emission is then the intersection of the
past light cone of the observer with this hypersurface.  We assume that
at conformal time $\ee$ every point with spatial coordinates
$p^i$ emits thermal radiation with a temperature
$T_\E(\p,\dh)$, as measured by a comoving observer,
which depends both on position and on direction as
characterized by a three-vector $\dh$, normalized to unity in the
background metric, $g^{(0)}_{\mu\nu}$,
restricted to the hypersurface. (This hypersurface
need not be the actual time of last scattering, but need only
represent a hypersurface on which the radiation field is
understood.) The photon path itself is specified by
a three-vector $\eh$ in the hypersurface of
constant conformal time containing $\O$ normalized
to unity in $g^{(0)}_{\mu\nu}$. We can think of $\eh$ as the
direction on the sky toward which a comoving observer at $\O$
is pointing an antenna; for observers which are not comoving
$\eh$ and the observer's direction vector are related by a Lorentz
transformation. The initial condition
$\eh$ determines the point $\p$ and
direction vector $\dh$ at which the ray intersects the
hypersurface of emission.

To an observer with four-velocity $U^\mu$ (normalized to
$U^\mu {\bar g}_{\mu\nu}U^\nu = -1$), a photon with wavevector
$k^\nu=dx^\nu/d\lambda$, with $\lambda$ an affine
parameter in the conformal metric $g_{\mu\nu}$,
has a relative frequency given by
  $$\omega = -a^{-2}{\bar g}_{\mu\nu} U^\mu k^\nu\ .\eqno(2.2)$$
(We refer to this as the ``relative'' frequency, since we are free
to scale the affine parameter $\lambda$ to set the normalization
of $\omega$.  The ratio of relative frequencies at two points along
the path is invariant under such a reparameterization.)
For a blackbody spectrum, the CMB temperature observed
at $\O$ is related
to the temperature at emission by
  $$T_\O(\x_\O,\eh) = {{\omega_\O}\over{\omega_\E}}T_\E(\p,\dh)\ .
  \eqno(2.3)$$
We are therefore interested in computing, given the initial
data $x^\mu_\O, \eh, \omega_\O$, the quantities $\p, \dh$,
and $\omega_\E$.  These depend on the photon path and associated
wavevector, which we may express as series expansions in the
perturbation $h_{\mu\nu}$ and its derivatives:
  $$\eqalign{x^\mu(\lambda) &= x^{(0)\mu}(\lambda)
  +x^{(1)\mu}(\lambda)+x^{(2)\mu}(\lambda)+\ldots \cr
  k^\mu(\lambda) &= k^{(0)\mu}(\lambda)
  +k^{(1)\mu}(\lambda)+k^{(2)\mu}(\lambda)+\ldots \cr}\eqno(2.4)$$
The situation is thus as portrayed in Fig.~1.  Note that
$x^{(0)\mu}(\lambda)$ has the interpretation of a path through
spacetime, while the $x^{(a)\mu}(\lambda)$
are thought of as deviation vectors at each $\lambda$.  In this
section we will calculate the observed temperature in terms of
these quantities (plus the intrinsic temperature fluctuations
on the surface of emission), while in the next section we will
explicitly calculate the path and wavevector in terms of the
metric perturbation.

We have already specified $T_{\E}$ as the temperature measured
by a comoving observer. It will also be convenient to take our
observer at $\O$ comoving.
This requirement is physically acceptable, since any motion of the
observer leads to a dipole anisotropy which may be easily subtracted.
It is sometimes useful to imagine a family of comoving observers
with four-velocity $U^{\mu}$ defined over all of spacetime.
The normalization condition $U^\mu {\bar g}_{\mu\nu}U^\nu = -1$
then leads to
  $$\eqalign{U^{\0\mu}&=a^{-1}(1,0,0,0)\cr
  U^{\1\mu}&=a^{-1}\left({1\over 2}h_{00},0,0,0\right)\cr
  U^{\2\mu}&=a^{-1}\left({3\over 8}(h_{00})^2,0,0,0\right)\
  .\cr}\eqno(2.5)$$
We can also explicitly construct the geodesics of the background
metric, $x^{\0\mu}(\lambda)$.  We consider
null rays which intersect the observer at the spatial origin of
co-ordinates, and we choose the affine parameter such that
  $$\eqalign{k^{\0 0}&=1\ ,\cr
  g^\0_{ij}k^{\0 i}k^{\0 j} &= 1\ .\cr}\eqno(2.6)$$
A two parameter family of such rays which satisfy these conditions
is given by [21]
  $$\eqalign{x^{(0)\mu}&=(\lambda, re^i)\cr
  k^{(0)\mu}&=(1, -\gamma e^i)\ ,\cr}\eqno(2.7)$$
where the $e^i$ are components of $\eh$, and
  $$\eqalign{r (\lambda )    &=2\tan_{\kappa }\left(
  {\lambda_{\O}-\lambda \over 2}\right)  \cr
          \gamma (\lambda ) &=\sec_{\kappa }^2\left(
  {\lambda_{\O}-\lambda \over 2}\right)\ , \cr} \eqno{(2.8)} $$
where $\lambda_{\O}$ is the affine
parameter at the observer. The subscript $\kappa$ on a
trigonometric function denotes a set of three functions:
for $\kappa =1$ the trigonometric function itself,
for $\kappa =-1$ the corresponding hyperbolic
function, and for $\kappa =0$ the first term
in the series expansion of the function.  (Thus, $\sin_0\theta
=\theta$, $\cos_0\theta = 1$.)  Finally, we can place boundary
conditions on the higher-order quantities $x^{(1)\mu}$,
$x^{(2)\mu}$, $k^{(1)\mu}$ and $k^{(2)\mu}$ at the origin.
For convenience we will set
  $$\eqalign{x^{(1)\mu}\left(\lambda_{\O}\right)
  =x^{(2)\mu}\left( \lambda_{\O}\right)&=0\cr
  k^{(1)i}\left(\lambda_{\O}\right)=k^{(2)i}
  \left(\lambda_{\O}\right)&=0\ .\cr}\eqno(2.9)$$
Then the condition that the
wavevector be null at the observer implies that
  $$\eqalign{k^{\1 0}(\lambda_\O)=& \left({1\over 2}h_{00} +
  h_{0i}k^{\0 i}+{1\over 2}h_{ij}k^{\0 i}k^{\0 j}\right)_\O \ ,\cr
  k^{\2 0}(\lambda_\O)=&\Biggl[{3\over 8}(h_{00})^2 +
  h_{00}h_{0i}k^{\0 i} + {1\over 4}h_{00}h_{ij}k^{\0 i}k^{\0 j}
  +{1\over 2}(h_{0i}k^{\0 i})^2 \cr
  &\qquad - {1\over 8}(h_{ij}k^{\0 i}
  k^{\0 j})^2\Biggl]_\O\ .\cr}\eqno(2.10)$$

The temperature at emission can be written as a uniform
background plus a small perturbation, expressed
as
  $$T_\E(\p, \dh) = [1+\tau(\p,\dh)]T^{(0)}_\E \ .\eqno(2.11)$$
The function $\tau$ will be treated as first order
({\it i.e.} of the same order as $h_{\mu\nu}$), and
will be unspecified in this paper
since our interest is in the gravitational effects on photons in the
time since emission.  The point at which the geodesic intersects the
surface $\eta=\eta_\E$ can be written as $\p=\p^{\0}+\p^{\1}+\ldots$
(Note the distinction between $x^i$,
the spacelike components of the separation vector, and
$p^i$, the separation of the intersection points of the path at
different orders with the constant-time hypersurface.)  Expanding
$\omega_\E$ and $\dh$ as well, eq.~(2.3) to second order becomes
  $$T_\O = {{\omega_\O^\0+\omega_\O^\1+\omega_\O^\2}\over
  {\omega_\E^\0+\omega_\E^\1+\omega_\E^\2}}
  [1+\tau(\p^\0+\p^\1,\dh^\0+\dh^\1)]T^\0_\E\ . \eqno(2.12)$$
With the conventions chosen in the previous paragraph,
$\omega^\0_\O=a\left( \eta_{\O}\right)^{-1}$ and
$\omega^\0_\E=a\left( \eta_{\E}\right)^{-1}$.
The quantity of interest to us is the fractional deviation in the
observed temperature with respect to the expected temperature in
the unperturbed spacetime, and we denote this deviation by $\dT$.
Expanding $\tau$ in a Taylor series, we obtain
  $$\eqalign{\dT &\equiv\left({{\omega_\E^\0}\over
   {\omega^\0_\O}}\right)
  {{T_\O}\over{T_\E^\0}}\cr &=
  \Biggl[ 1+\left({\tilde \omega}^\1_\O -{\tilde \omega}^\1_\E
  +\tau\right)\cr
  &~~+\left({\tilde \omega}^\2_\O - {\tilde \omega}^\2_\E
  +({\tilde \omega}^\1_\E)^2
  -{\tilde \omega}^\1_\O{\tilde \omega}^\1_\E
  +{\tilde \omega}^\1_\O \tau
  -{\tilde \omega}^\1_\E \tau+
   p^{\1 i}{{\partial\tau}\over{\partial x^i}}
  + d^{\1 i}{{\partial\tau}\over{\partial d^i}}
  \right)\Biggl]\ ,\cr}\eqno(2.13)$$
where the $d^i$ are the components of $\dh$, $\tau$ and
its first partial derivatives are
evaluated at $(\p^\0, \dh^\0)$, and we have put
${\tilde \omega}^{(a)}=\omega^{(a)}/\omega^{(0)}$. We note that
our freedom to choose $T_\E^\0$ may be used to render
$\dT$ observable, e.g. by setting $T_\E^\0=a\left(
\eta_{\O}\right)a\left( \eta_{\E}\right)^{-1}\langle T_{\O}
\rangle $ where the angle brackets denote an average over the
observer's sky.

Expanding the metric perturbation and photon wavevector around their
values on the background path, we obtain
  $$\eqalign{{\tilde \omega}^\0 &= 1\cr
  {\tilde \omega}^\1 &= -{1\over 2}h_{00}
  - k^{\0 i}h_{0i} + k^{\1 0}\cr
  {\tilde \omega}^\2 &= -{1\over 8}(h_{00})^2
  - {1\over 2}h_{00} k^{\1 0}
  - {1\over 2}k^{\0 i} h_{0i} h_{00} - h_{0i} k^{\1 i} + k^{\2 0}
  \cr & \qquad
  -{1\over 2}p^{\1 i}{{\partial h_{00}}\over{\partial x^i}}
  - k^{\0 i} p^{\1 j}{{\partial h_{0i}}\over{\partial x^j}}
  +\Delta\lambda{{dk^{\1 0}}\over{d\lambda}}
  -h_{0i}\Delta\lambda{{dk^{\0 i}}\over{d\lambda}}\ .\cr}\eqno(2.14)$$
In this expression $\Delta\lambda$ is the difference in affine
parameter between the point where the zeroth and first order
geodesics intersect the hypersurface $\eta=$~constant; to this
order $\Delta\lambda = -x^{\1 0}$.  It is also straightforward to show
that $p^{\1 i} = x^{\1 i} -k^{\0 i} x^{\1 0}$, and that $d^{\1 i}$
is given by
  $$d^{\1 i} = {{k^{\0 i}+k^{\1 i}}\over{|k^{\0 i}+k^{\1 i}|}}
  - {{k^{\0 i}}\over{|k^{\0 i}|}}\ ,\eqno(2.15)$$
where the norm is defined by the spacelike part of the
background metric.  Putting it all together we obtain
  $$\eqalign{\dT^\0 = & 1\cr
  \dT^\1 = & \left[{1\over 2}
  h_{ij}k^{\0 i}k^{\0 j}\right]_\O +
  \left[{1\over 2}h_{00} +h_{0i}k^{\0 i} - k^{\1 0}
  +\tau\right]_{\Eb}\cr
  \dT^\2 =& \Biggl[{1\over 2}(h_{0i}k^{\0 i})^2
  -{1\over 8}(h_{ij}k^{\0 i}k^{\0 j})^2\Biggl]_\O
  \cr &\
  +\Biggl[{1\over 2}h_{ij}k^{\0 i}k^{\0 j}\Biggl]_\O
  \Biggl[{1\over 2}h_{00}+ h_{0i}k^{\0 i}-k^{\1 0}+\tau\Biggl]_{\Eb}
  \cr &\
  +\Biggl[{3\over 8}(h_{00})^2 - {1\over 2}h_{00} k^{\1 0}
  + {3\over 2} h_{0i} h_{00}k^{\0 i} + (h_{0i}k^{\0 i})^2
  -2h_{0i}k^{\0 i}k^{\1 0}
  \cr &\ \
  + h_{0i} k^{\1 i}+ (k^{\1 0})^2 - k^{\2 0}
  +({1\over 2}h_{00} + h_{0i}k^{\0 i} - k^{\1 0})\tau
  +x^{\1 0}{{dk^{\1 0}}\over{d\lambda}}
  \cr &\ \
  -h_{0i}x^{\1 0}{{dk^{\0 i}}\over{d\lambda}}
  +(x^{\1 i}-k^{\0 i} x^{\1 0})\left({1\over 2}
  {{\partial h_{00}}\over{\partial x^i}}
  +k^{\0 j} {{\partial h_{0j}}\over{\partial x^i}}
  + {{\partial\tau}\over{\partial x^i}}\right)
  + d^{\1 i}{{\partial\tau}\over{\partial d^i}}\Biggl]_{\Eb}.
  \cr}\eqno(2.16)$$
Here, the notation $\Eb$ means that the quantities referred to should
be evaluated at the point $(\eta_\E,\p^\0)$ and direction $\dh^\0$.

To complete the above formulae, we have to solve for the perturbed
geodesics at first and second order in terms of $h_{\mu\nu}$.
In the next section we carry this out for arbitrary metric
perturbations, and in the following section we specialize to scalar
perturbations.

\vskip .1in
{\bf III. Second-Order Geodesics}

\nobreak
In order to calculate the approximate geodesics of $g_{\mu\nu}=
g^\0_{\mu\nu} +h_{\mu\nu}$ order by order we employ
the perturbative geodesic expansion
introduced in Pyne and Birkinshaw [20]. Because those authors
worked only to first order it is necessary slightly to extend the
equations to address the higher order questions we are
concerned with here. In this section we describe the
needed extension, which writes a general solution for the
approximate path at any order without restriction on the perturbed
spacetime under consideration. In the following section we specialize
this general solution to gain the null geodesics
to second order of perturbed FRW spacetimes in the longitudinal gauge.

We begin with the geodesic
equation in the metric $g_{\mu\nu}=g^{(0)}_{\mu\nu}+h_{\mu\nu}$,
  $${d^2 x^{\mu}\over d\lambda^2}+\Gamma^{\mu}{}_{\alpha\beta}
k^{\alpha}k^{\beta}=0\ , \eqno{(3.1)}$$
which holds along some path $x^{\mu}(\lambda )$. We seek to approximate
that path to any given order by solving for the $x^{(a)\mu}(\lambda )$
in (2.4). To this end we substitute (2.4) and the equation
  $$\Gamma^{\mu}{}_{\alpha\beta}=\Gamma^{(0)\mu}{}_{\alpha\beta}
  +\Gamma^{(1)\mu}{}_{\alpha\beta}+
  \Gamma^{(2)\mu}{}_{\alpha\beta}+...\eqno{(3.2)}$$
into (3.1) and simultaneously
Taylor expand each of the
$\Gamma^{(a)\mu}{}_{\alpha\beta}$ at $x^{\mu}(\lambda )$
about their value at $x^{(0)\mu}(\lambda )$. In (3.2),
$\Gamma^{(a)\mu}{}_{\alpha\beta}$ is that part of
$\Gamma^{\mu}{}_{\alpha\beta}$ which is of $a$-th order in either
$h_{\mu\nu}$, its first partial derivatives, or their products.
The resulting
equation, equivalent to (3.1) but holding along the path
$x^{(0)\mu}(\lambda )$, is written
  $$\eqalign{ \Sigma_{a=0}^{\infty}\Biggl[
  {d^2 x^{(a)\mu}\over d\lambda^2}
  &+\left( \Gamma^{(a)\mu}{}_{\alpha\beta} +\Sigma_{b=1}^{\infty} {1\over b!}
   \partial_{\sigma_1}\cdots \partial_{\sigma_b}
  \Gamma^{(a)\mu}{}_{\alpha\beta}\left( \Sigma_{c=1}^{\infty}
  x^{(c)\sigma_1}\right)\cdots \left( \Sigma_{d=1}^{\infty}
  x^{(d)\sigma_l}\right) \right) \cr
  &\qquad \times\left( \Sigma_{e=1}^{\infty}
  k^{(e)\alpha}\right)\left( \Sigma_{f=1}^{\infty}
  k^{(f)\beta}\right) \Biggr] =0\cr}\eqno{(3.3)}$$
At zeroth order we find that $x^{(0)\mu}(\lambda )$ is an affinely
parametrized geodesic in the metric $g^{(0)}_{\mu\nu}$.
At every order above zeroth equation (3.3) may be rearranged into the form of
a forced Jacobi equation for the
$a$-th order separation vector, $x^{(a)\mu}(\lambda )$:
  $${d^2 x^{(a)\mu}\over d\lambda^2}+2\Gamma^{(0)\mu}{}_{\alpha\beta}
  k^{(0)\alpha}k^{(a)\beta}+\partial_\sigma\Gamma^{(0)\mu}{}_{\alpha\beta}
  k^{(0)\alpha}k^{(0)\beta}x^{(a)\sigma}=f^{(a)\mu} \ .\eqno{(3.4)}$$
Importantly, the highest order $x^{(b)\mu}$ or
$k^{(b)\mu}$ appearing
in $f^{(a)\mu}$ is of ($a-1$)-th order.
For instance, the forcing vectors at first and second order are
given by
  $$\eqalign{ f^{(1)\mu}&=-\Gamma^{(1)\mu}{}_{\alpha\beta}
  k^{(0)\alpha}k^{(0)\beta} \cr
  f^{(2)\mu}&=-\Gamma^{(0)\mu}{}_{\alpha\beta}k^{(1)\alpha}k^{(1)\beta}
  -2\Gamma^{(1)\mu}{}_{\alpha\beta}k^{(0)\alpha}k^{(1)\beta}
  -2\partial_\sigma\Gamma^{(0)\mu}{}_{\alpha\beta}
  x^{(1)\sigma}k^{(0)\alpha}k^{(1)\beta} \cr
  &\qquad -\partial_\sigma\Gamma^{(1)\mu}{}_{\alpha\beta}
  x^{(1)\sigma}k^{(0)\alpha}k^{(0)\beta} -{1\over 2}
  \partial_\sigma\partial_\tau\Gamma^{(0)\mu}{}_{\alpha\beta}
  x^{(1)\sigma}x^{(1)\tau}k^{(0)\alpha}k^{(0)\beta} \cr
  &\qquad -
  \Gamma^{(2)\mu}{}_{\alpha\beta}k^{(0)\alpha}k^{(0)\beta}\ .\cr}
  \eqno{(3.5)}$$

Pyne and Birkinshaw [20] showed how a general solution to
equation (3.4) could be written down in terms of the parallel and Jacobi
propagators of the background metric $g^\0_{\mu\nu}$.  These
propagators are matrix-valued functions of a pair of points connected
by a geodesic, and are defined by path-ordered exponentials along
the appropriate geodesic.  The parallel propagator $P(\lambda_1,
\lambda_2)^\mu{}_\nu$ is a $4\times 4$ matrix given by
  $$P\left(\lambda_2 ,\lambda_1\right)={\cal P}\exp \left(
  -{1\over 2}\int_{\lambda_1}^{\lambda_2} A(\lambda)
  \ d\lambda \right)\ ,\eqno{(3.6)}$$
where $\cal P$ denotes the path ordering symbol and $A$ is a
$4\times 4$ matrix defined by $A^\mu{}_\nu = 2k^{\0 \sigma}
\Gamma^{\0\mu}{}_{\sigma\nu}$.  The parallel propagator
lives up to its name, in that $P\left( \lambda_1 ,\lambda_2\right)
^{\mu}{}_{\nu}v^{\nu}\left( \lambda_2\right)$ is the vector
obtained by parallel propagating $v^\mu$ from $\lambda_2$ to
$\lambda_1$ along the geodesic.  The Jacobi propagator is an
$8\times 8$ matrix given by
  $$U\left( \lambda_2,\lambda_1\right)={\cal P}\exp \left(
  \int_{\lambda_1}^{\lambda_2}\pmatrix{ 0 &1 \cr P\left(
  \lambda_1,\lambda\right){\cal R}\left(\lambda\right)P\left(\lambda ,
  \lambda_1\right) & 0\cr}d\lambda \right)  ,\eqno{(3.7)}$$
where ${\cal R}(\lambda )^{\mu}{}_{\sigma}$ denotes the
$4\times 4$ matrix $R^{(0)\mu}{}_{\nu\rho\sigma}
k^{(0)\nu}k^{(0)\rho}$ evaluated at $x^{(0)}(\lambda)$, and $0$ and
$1$ denote the $4\times 4$ zero and identity matrices, respectively.
The Jacobi propagator serves as a Green's function for the Jacobi
equation in the background spacetime.  More information about these
objects can be found in [20, 22].

The solution for $x^{(a)\mu}(\lambda )$ and $k^{(a)\mu}(\lambda)$
at some affine parameter $\lambda_2$ can now be obtained from
their values at some fixed affine parameter
$\lambda_1$ via
  $$\eqalign{ \pmatrix{ P\left( \lambda_1 ,\lambda_2 \right)
  x^{(a)}\left(
  \lambda_2\right) \cr {d\over d\lambda_2}\left[ P\left( \lambda_1 ,
  \lambda_2\right)x^{(a)}\left( \lambda_2\right) \right]}
   &=U\left( \lambda_2 ,\lambda_1\right)\pmatrix{ x^{(a)}\left(
  \lambda_1 \right) \cr \left[ {d\over d\lambda}
  \left[ P\left( \lambda_1 ,
  \lambda \right)x^{(a)}(\lambda )\right]\right]_{\lambda=\lambda_1} } \cr
   &\qquad +\int_{\lambda_1}^{\lambda_2}
   U\left( \lambda_2 ,\lambda \right)
  \pmatrix{ 0 \cr P\left( \lambda_1 ,\lambda \right)f^{(a)}(\lambda )}
  d\lambda\ , } \eqno{(3.8)}$$
the integral being taken over the zeroth order geodesic, $x^{(0)\mu}
(\lambda )$. The program for recursive calculation of the $x^{(a)\mu}
(\lambda )$ is now established: having obtained
$x^{(a-1)\mu}(\lambda )$ we can solve for $f^{(a)\mu}$ and thus
obtain $x^{(a)\mu}(\lambda )$ from (3.8). The recursion starts
by solving the geodesic equation of the background for some
$x^{(0)\mu}(\lambda )$ and then calculating its associated parallel
and Jacobi propagators.

The parallel and Jacobi propagators for the radial, null
geodesics of $g^{(0)}_{\mu\nu}$, the conformally transformed
Robertson-Walker metric, were obtained in [23] and are written
  $$P\left( \lambda_2 ,\lambda_1\right)^{\mu}{}_{\nu}=
  \pmatrix{ 1 &0_j \cr 0^i & {\gamma\left( \lambda_2 \right)\over
  \gamma\left( \lambda_1 \right)}\delta^i_j \cr}  \eqno{(3.9)}$$
and
  $$U\left( \lambda_2 ,\lambda_1 \right) =
  \pmatrix{ \cos_{\kappa}\left( \lambda_2 -\lambda_1\right)J&
  \sin_{\kappa}\left( \lambda_2 -\lambda_1\right)J \cr
  -\kappa \sin_{\kappa}\left( \lambda_2 -\lambda_1\right)J &
  \cos_{\kappa}\left( \lambda_2 -\lambda_1\right)J& \cr}
  +\pmatrix{ (1 -J) & \left(
  \lambda_2 -\lambda_1\right)(1 -J) \cr 0 & (1 -J) \cr}
  \eqno{(3.10)}$$
respectively. In (3.10) $J$ is a $4\times 4$ matrix given by
  $$J^{\mu}{}_{\sigma}=\pmatrix{ 0 & 0_j \cr 0^i & \delta^i_j
  -e^ie_j \cr}\ . \eqno{(3.11)}$$
Given a background geodesic specified by the direction cosines $e^i$,
any three-vector $v^i(\lambda)$
may be decomposed into the sum of a longitudinal part
$v^i_\Vert(\lambda)$ pointing along the geodesic and a transverse
part $v^i_\perp(\lambda)$ perpendicular to the geodesic (in the spacelike
hypersurface of the background), where
  $$\eqalign{v^i_\Vert(\lambda) &= e^i e_j v^j(\lambda)\ ,\cr
  v^i_\perp(\lambda) &= (\delta^i_j-e^i e_j)v^j(\lambda)\ .\cr}
  \eqno{(3.12)}$$
Thus, the matrix $J$ serves to project a four-vector into
the plane transverse to the photon direction in the
comoving spatial hypersurfaces.

The relatively simple form of these propagators for the case
of a Robertson-Walker metric allows us to obtain the perturbed
geodesic and wavevector from (3.8) immediately.  Imposing the
boundary conditions (2.9) at the observer, we obtain
  $$\eqalign{x^{(a)0}(\lambda) & = (\lambda-\lambda_\O)
  k^{(a)0}(\lambda_\O) + \int_{\lambda_\O}^\lambda
  (\lambda-\lambda')f^{(a)0}(\lambda') d\lambda'\ ,\cr
  x^{(a)i}_\Vert(\lambda) & = \gamma(\lambda)\int_{\lambda_\O}^\lambda
  (\lambda-\lambda')\gamma^{-1}(\lambda')f^{(a)i}_\Vert(\lambda')
  d\lambda' \ ,\cr
  x^{(a)i}_\perp(\lambda) & = \gamma(\lambda)\int_{\lambda_\O}^\lambda
  \sin_\kappa(\lambda-\lambda')\gamma^{-1}(\lambda')f^{(a)i}_\perp(\lambda')
  d\lambda' \ ,\cr
  k^{(a)0}(\lambda) & = k^{(a)0}(\lambda_\O) +\int_{\lambda_\O}^\lambda
  f^{(a)0}(\lambda')d\lambda' \ ,\cr
  k^{(a)i}_\Vert(\lambda) & = \gamma(\lambda)\int_{\lambda_\O}^\lambda
  \left[1-{{\kappa r(\lambda)}\over 2}(\lambda-\lambda')\right]
  \gamma^{-1}(\lambda')f^{(a)i}_\Vert(\lambda') d\lambda' \ ,\cr
  k^{(a)i}_\perp(\lambda) & = \gamma(\lambda)\int_{\lambda_\O}^\lambda
  \left[\cos_\kappa(\lambda-\lambda')-{{\kappa r(\lambda)}\over 2}
  \sin_\kappa(\lambda-\lambda')\right]\gamma^{-1}(\lambda')
  f^{(a)i}_\perp(\lambda') d\lambda' \ .\cr}\eqno(3.13)$$
These expressions are valid for any metric perturbation; the specific
perturbation is encoded in the vectors $f^{(a)\mu}$.  Note that
we have written the integrals as proceeding backwards along the path
from the observer to the point $\lambda$ on the background path.

\vskip .1in
\noindent{\bf IV. Scalar Perturbations in the Longitudinal Gauge}

\nobreak
In this section we carry out the program
described above to second order
for the metric perturbation $h_{\mu\nu}$ given by
  $$h_{\mu\nu}dx^{\mu}dx^{\nu}=-2\phi d\eta^2-2\psi\gamma^{-2}\left(
  dx^2+dy^2+dz^2\right) \eqno{(4.1)}$$
describing scalar perturbations in the longitudinal
gauge [24]. We note that
in this gauge, $\phi$ and $\psi$ coincide with the
gauge invariant metric variables of [24], $\Phi$ and $\Psi$.
This will allow us to obtain gauge invariant expressions for the
observables of interest by replacing $\phi$ with $\Phi$ and $\psi$
with $\Psi$ in our final formulae. Of course, only the first order
expressions are rendered gauge invariant because $\Phi$ and $\Psi$
are themselves gauge invariant only to first order.

To compute the perturbation vectors $f^{(a) \mu}$ we
need to calculate the Christoffel symbols to various orders.  These
are given by
  $$\eqalign{\G^{\0 0}{}_{\sigma\alpha} &= 0 \cr
  \G^{\0 i}{}_{0\alpha} &= 0 \cr
  \G^{\0 i}{}_{jk} &= -{{\kappa}\over{2\gamma}}(\delta_{ik} x^\0_j
   +\delta_{ij} x^\0_k-\delta_{jk} x^\0_i) \cr
  \G^{\1 0}{}_{0\alpha} &= \partial_\alpha\phi \cr
  \G^{\1 0}{}_{ij} &= -{{\partial_0\psi}\over{\gamma^2}}\delta_{ij} \cr
  \G^{\1 i}{}_{00} &= \gamma^2\partial_i\phi \cr
  \G^{\1 i}{}_{0j} &= -\partial_0\psi\delta_{ij} \cr
  \G^{\1 i}{}_{jk} &= -\delta_{ik} \partial_j\psi
   -\delta_{ij} \partial_k\psi+\delta_{jk}\partial_i\psi  \cr
  \G^{\2 0}{}_{0\alpha} &= -2\phi\partial_\alpha\phi \cr
  \G^{\2 0}{}_{ij} &= {{2\phi\partial_0\psi}\over{\gamma^2}}\delta_{ij} \cr
  \G^{\2 i}{}_{00} &= 2\gamma^2\psi\partial_i\phi \cr
  \G^{\2 i}{}_{0j} &= -2\psi\partial_0\psi\delta_{ij} \cr
  \G^{\2 i}{}_{jk} &= -2\psi(\delta_{ik} \partial_j\psi
   +\delta_{ij} \partial_k\psi-\delta_{jk}\partial_i\psi)\ .  \cr}
  \eqno(4.2)$$

Calculation of the first order vector $f^{\1\mu}$ proceeds
straightforwardly, using the normalization
$g^{\0}_{ij}k^{\0 i}k^{\0 j}=\delta_{ij}e^ie^j=1$.
We find that
  $$\eqalign{f^{\1 0} &= \partial_0(\phi+\psi) -
  2{{d\phi}\over{d\lambda}}\ ,\cr
  f^{\1 i}_\Vert &= -k^{\0 i}k^{\0 j}\partial_j(\phi+\psi)
  +2 k^{\0 i}{{d\psi}\over{d\lambda}}\ ,\cr
  f^{\1 i}_\perp &= (k^{\0 i}k^{\0 j}-g^{\0 ij})
  \partial_j(\phi+\psi) \ .\cr}\eqno(4.3)$$
According to (2.16), the only second-order quantity (as distinguished
from products of first-order quantitites) which enters
the formula for $\dT^\2$ is the timelike component of the
wavevector, $k^{\2 0}$.  We therefore do not need to calculate the
entire second-order force vector, but only the timelike component.
In doing so we make use of the decomposition of the directional
derivative of a scalar along the path into partial derivatives,
  $${{d\phi}\over{d\lambda}} = \partial_0\phi + k^{\0 i}\partial_i\phi
  \eqno(4.4)$$
(since $k^{\0 0}=1$).  Another relatively straightforward calculation
yields
  $$\eqalign{f^{\2 0}= -&2{d\over{d\lambda}}\left(k^{\1 0}\phi +
  x^{\1\sigma}\partial_\sigma\phi\right) +2k^{\1 0}
  \partial_0(\phi-\psi)\cr  &+ x^{\1\sigma}\partial_\sigma\partial_0
  (\phi+\psi) +2(\phi+\psi)\partial_0\psi\ .\cr}\eqno(4.5)$$

Substituting (4.3) into (3.13), we obtain the first order perturbed
geodesic:
  $$\eqalign{x^{(1)0}\left( \lambda \right) &=\left(\lambda -
  \lambda_{\O}\right)
  \left( \phi-\psi\right)_\O +\int_{\lambda_{\O}}^{\lambda}
  \left[-2\phi+\left(\lambda -\lambda' \right)\partial_0(\psi+\phi)
  \right]\, d\lambda' \ ,\cr
  x^{(1)i}_\Vert(\lambda) & = (\lambda - \lambda_\O)(\phi-\psi)_\O
  k^{\0 i}(\lambda) + k^{\0 i}(\lambda)\int_{\lambda_{\O}}^{\lambda}
  \left[(\psi-\phi)+\left(\lambda -\lambda' \right)\partial_0(\psi+\phi)
  \right]\, d\lambda' \ ,\cr
  x^{(1)i}_\perp(\lambda) & = \gamma(\lambda)
  \int_{\lambda_{\O}}^{\lambda}
  \sin_{\kappa } \left( \lambda -\lambda'\right)
  \gamma(\lambda')\left[e^ie^j-\delta^{ij}\right]
  \partial_j(\phi +\psi)\, d\lambda'\ .\cr}\eqno{(4.6)}$$

The explicit construction for the wavevector perturbation,
$k^{(1)}(\lambda )$, may be obtained either by differentiation of
(4.6) above or from (3.13) directly. In either case
  $$\eqalign{k^{(1)0}\left( \lambda\right) & =
  \left(\phi-\psi\right)_{\O} -2\phi_\lambda - I_{\rm ISW}(\lambda)\ ,\cr
  k^{(1)i}_\Vert\left(\lambda\right)
     &=-{{\kappa r(\lambda)}\over 2}x^{(1)i}_\Vert
  \left( \lambda\right) +k^{(0)i}\left(\lambda\right)
  \left[(\phi-\psi)_\O - (\phi-\psi)_\lambda - I_{\rm ISW}(\lambda)
  \right]\ ,\cr
  k^{(1)i}_\perp\left(\lambda\right) & = \gamma(\lambda)
  \int_{\lambda_\O}^\lambda
  \left[\cos_\kappa(\lambda-\lambda')-{{\kappa r(\lambda)}\over 2}
  \sin_\kappa(\lambda-\lambda')\right]\gamma(\lambda')
  \left[e^ie^j-\delta^{ij}\right]
  \partial_j(\phi +\psi) d\lambda' \ ,\cr}\eqno(4.7)$$
where the integral
  $$I_{\rm ISW}(\lambda) = -\int_{\lambda_{\O}}^{\lambda}
  \partial_0(\psi+\phi) d\lambda'\eqno(4.8)$$
represents the conventional integrated Sachs-Wolfe effect.

As noted above, to compute the second-order effect we need only
the time component of the second-order wavevector; this is given
by
  $$k^{\2 0}(\lambda) =-{1\over 2}(\phi+\psi)^2_\O
  -2(x^{\1\mu}\partial_\mu\phi +k^{\1 0}\phi)_\lambda
  +I_2(\lambda)\ ,\eqno(4.9)$$
where the integral $I_2$ is defined as
  $$I_2(\lambda) = \int_{\lambda_\O}^{\lambda}\left[2k^{\1 0}
  \partial_0(\phi-\psi)+2(\phi+\psi)\partial_0\psi+x^{\1\mu}
  \partial_\mu\partial_0(\phi+\psi)\right]d\lambda'\ .\eqno(4.10)$$

Having obtained the perturbed geodesics and
wavevectors in the longitudinal gauge, it remains only to substitute
into (2.16) to obtain final expressions for the temperature
anisotropy.  At first order we recover the conventional
Sachs-Wolfe result,
  $$\dT^\1 = (\phi + \tau)_{\Eb} -\phi_\O +I_{\rm ISW}(\lambda_{\Eb})
  \ ,\eqno(4.11)$$
where once again the notation ${\Eb}$ means that quantities are
evaluated at the position and direction of the intersection of the
background geodesic with the surface of emission.
The second order anisotropy, the main result of this paper,
is given by
  $$\eqalign{\dT^\2 = & {3\over 2}\phi^2_\O - \phi_\O\phi_{\Eb}
  -\phi_\O\tau_{\Eb} - {1\over 2}\phi^2_{\Eb} + \phi_{\Eb}\tau_{\Eb} \cr
  & -\left(2\phi_\O -\psi_\O - \phi_{\Eb} -\tau_{\Eb}
  - I_{\rm ISW}(\lambda_{\Eb})\right)I_{\rm ISW}(\lambda_{\Eb})
  - I_2(\lambda_{\Eb}) \cr
  & +\left(x^{\1 i}_\perp
  +I_{\rm TD} k^{\0 i}\right)_{\Eb}\partial_i(\phi+\tau)_{\Eb}
  + x^{\1 0}\partial_0(\phi+\psi)_{\Eb}
  + d^{\1 i}{{\partial\tau}\over{\partial d^i}}\ ,\cr}\eqno(4.12)$$
where the integral
  $$I_{\rm TD}(\lambda) = \int_{\lambda_\O}^{\lambda}(\phi+\psi)
  d\lambda'\eqno(4.13)$$
is the Shapiro time delay along the path.

An accurate appraisal of the magnitudes of the various terms
contained in (4.12) would require knowledge of the initial conditions
and evolution of the perturbations $\phi$ and $\psi$, including
nonlinear effects.  This information is model-dependent,
and we will not attempt such a task here.  It is nevertheless
possible to remark on the possible importance of the different
effects to observations, based simply on the form in which they
appear.

The quantities $\phi_\O$, $\psi_\O$, $\phi_{\Eb}$, $\tau_{\Eb}$ and
$I_{\rm ISW}(\lambda_{\Eb})$ are all small ($\leq 10^{-5}$)
in conventional models
of structure formation.  Therefore the terms in (4.12) which are
written as products of these numbers are even smaller, and should
not contribute to the anisotropy at an observable level.
Similarly the term
$d^{\1 i}(\partial\tau/\partial d^i)$ will typically
be the product of two small quantities, and may be neglected.
Therefore the potentially interesting terms are those involving
the separation vector $x^{\1 \mu}(\lambda_{\Eb})$ (which is not
necessarily small) and the integrals $I_{\rm TD}(\lambda_{\Eb})$
and $I_2(\lambda_{\Eb})$.

The term $x^{\1 i}_\perp\partial_i(\phi+\tau)_{\Eb}$ is due
to the transverse deflection of the photons by sources
between us and the surface of emission; this effect has been
studied previously in investigation of the impact of
gravitational lenses on CMB anisotropy [25-36].  (The processing
of CMB anisotropy by lensing is second order since both the
lens angle and the initial fluctuations being processed are
themselves first order in our accounting scheme.)  While the
effect of lensing on the CMB perturbation spectrum has been
somewhat controversial, it can play an observable
role on small angular scales.  The fact that this second-order
effect may be significant can be thought of as a consequence
of the fact, noted in the Introduction, that the existence of
large distance scales in the problem can enhance higher-order
effects; in this case the transverse deflection, given
approximately by the product of the distance travelled times
the lens angle, builds up as the photon travels along its
trajectory.

The term $I_{\rm TD} k^{\0 i}\partial_i(\phi+\tau)_{\Eb}$ is the
longitudinal equivalent of the transverse lensing term.  It
arises from the time delay effect of the lenses, which alters
the spacelike distance between the observer and the point
where the photon path intersects the surface of emission.
The qualitative effect of this term is similar to that of the
transverse lensing term, although its magnitude is expected to
be smaller; for typical lens systems, the longitudinal
deflection is smaller than the transverse deflection by a
factor proportional to the lens angle ({\it i.e.}, by several
orders of magnitude).

The term $x^{\1 0}\partial_0(\phi+\psi)_{\Eb}$ arises because the
difference in affine parameter between observer and surface of
last scattering differs along the true and background paths.
It is similar in structure to the effects discussed in the
previous two paragraphs, but is presumably smaller since
the time derivatives of the potentials are typically smaller than
the spatial derivatives.

Finally the integral $I_2(\lambda_{\Eb})$ contains three terms.
The first two terms appear small; they are integrals of products
of two small quantities, and furthermore contain time derivatives
which are typically suppressed with respect to spatial derivatives.
The third term represents a correction to the ISW effect, taking into
account that the perturbations along the first-order path differ from
those along the background path.  In cold dark matter models with
$\Omega=1$ and adiabatic density perturbations, the
ISW effect itself is smaller than the conventional Sachs-Wolfe
term [8, 37-48], and the correction described here is presumably smaller
still; nevertheless, it is possible that observations of the CMB
will reach a level of precision at which this term should be taken
into account.  Moreover, time derivatives and the ISW effect can
be important in models of structure formation based on
topological defects [12-15], open universe models [49-54], and
models with an appreciable cosmological constant [55, 56].
In these types of universes the new terms represented by
$I_2(\lambda_{\Eb})$ could play a role analogous to that of
gravitational lensing in adiabatic CDM models.

\vskip .1in
\noindent{\bf V. Conclusions}

\nobreak
We have computed the anisotropy induced in the cosmic microwave
background, due to gravitational effects, to second order in a
given metric perturbation.  For an arbitrary perturbation, our
results are given by the basic equation (2.16) plus the solutions
(3.13) for the perturbed geodesic and wavevector, where the forcing
vectors to first and second order are given by (3.5). In the case
of scalar perturbations in the longitudinal gauge, these results
may be combined into the single compact formula (4.12).

Our results are reassuring for studies to date of CMB anisotropy,
in that they do not reveal any new effects which are likely to
dominate the anisotropy spectrum on any scale.  An informal
examination of our final expressions indicates that the effect
most likely to be observable is that due to (transverse)
gravitational lensing, which has already been the subject of some
attention in the literature.  As both theoretical and observational
studies of the CMB increase in accuracy and sophistication,
however, we feel it is important to know the precise form of
the effects we have explored.

With the basic framework in hand, there is clearly room for
future work along these lines.  One direction would be to investigate
a wider class of perturbations ({\it i.e.}, vector and tensor
modes), as well as to study carefully the second-order metric
perturbation itself.  An equally important task is to examine
the effects we have described more quantitatively, in the context
of a specific and detailed model of structure formation;
only then could we be completely confident in our understanding of
the role played by second-order perturbations in CMB anisotropy.

\vskip .1in
\noindent{\bf Acknowledgments}

We would like to thank Mark Birkinshaw, Arthur Kosowsky and Uros
Seljak for very helpful comments, and the referee for useful
suggestions.  This work was supported in part by
the National Science Foundation under grants AST/90-05038 and
PHY/9200687, by NASA contract NAS8-39073, and by the U.S. Department
of Energy (D.O.E.) under cooperative agreement DE-FC02-94ER40818.

\vfill\eject

\centerline{\bf References}

\item{[1]} M. White, D. Scott and J. Silk, {\it Ann. Rev. Astron.
Astrophys.} {\bf 32}, 319 (1994).

\item{[2]} J.R. Bond, in {\it Cosmology and Large Scale Structure},
Proceedings of Les Houches School, Summer 1993, ed. R. Schaefer,
Amsterdam:Elsevier (1995).

\item{[3]} U. Seljak, {\it Astrophys. J.} {\bf 435}, L87 (1994);
astro-ph/9406050.

\item{[4]} W. Hu and N. Sugiyama, {\it Phys. Rev. D}
{\bf 51} 2599 (1995); astro-ph/9411008.

\item{[5]} D. Scott, J. Silk and M. White, preprint CFPA-94-TH-51
(1995); astro-ph/9505015.

\item{[6]} R. K. Sachs and A. M. Wolfe, {\it Astrophys. J.}
{\bf 147}, 73 (1967)

\item{[7]} L. P. Grischuk and Y. B. Zeldovich,
{\it Sov. Astron.-AJ} {\bf 22}, 125 (1978).

\item{[8]} M. J. Rees and D. W. Sciama, {\it Nature} {\bf 217},
511 (1968).

\item{[9]} E. V. Linder,
{\it Astrophys. J.} {\bf 326}, 517 (1988).

\item{[10]} F. Argueso and E. Mart\'inez-Gonz\'alez,
{\it Mon. Not. Roy. Astron. Soc.} {\bf 238}, 1431 (1989).

\item{[11]}  F. Argueso, E. Mart\'inez-Gonz\'alez,
and J. L. Sanz, {\it Astrophys. J.}
{\bf 336}, 69 (1989).

\item{[12]} B. Allen, R.R. Caldwell, E.P.S. Shellard, A. Stebbins and
S. Veeraraghavan, preprint FERMILAB-CONF-94-197-A (1994); astro-ph/9407042.

\item{[13]} L. Perivolaropoulos, {\sl Astrophys. J.} {\bf 451}, 429 (1995);
astro-ph/9402024.

\item{[14]} R. G. Crittenden and N. Turok, preprint PUPT-1545 (1995);
astro-ph/9505120.

\item{[15]} R. Durrer, A. Gangui, and M. Sakellariadou, preprint
SISSA-83-95-A (1995); astro-ph/9507035.

\item{[16]} M. W. Jacobs, E. V. Linder, and R. V. Wagoner,
{\it Phys. Rev. D} {\bf 45}, R3292 (1992).

\item{[17]} E. Vishniac, {\it Astrophys. J.} {\bf 322}, 597 (1987).

\item{[18]} S. Dodelson and J. Jubas, {\it Astrophys. J.} {\bf 439},
503 (1995); astro-ph/9308019.

\item{[19]} W. Hu, D. Scott, and J. Silk, {\it Phys. Rev. D}
{\bf 49}, 648 (1994); astro-ph/9305038.

\item{[20]} T. Pyne and M. Birkinshaw, {\it Astrophys. J.}
{\bf 415}, 459 (1993); astro-ph/9303020.

\item{[21]} G. C. McVittie, {\it General Relativity and Cosmology},
2nd ed., Chapman and Hall, London (1964).

\item{[22]} J. L. Synge, {\it Relativity: The General Theory},
North-Holland, Amsterdam (1960).

\item{[23]}  T. Pyne and M. Birkinshaw, {\it Astrophys. J.} in press
(1995); astro-ph/9504060.

\item{[24]} V. F. Mukhanov, H. A. Feldman, and R. H. Brandenberger,
{\it Phys. Rep.} {\bf 215}, Nos. 5 \& 6, 203 (1992).

\item{[25]} A. Blanchard and J. Schneider, {\it Astron. Astrophys.}
{\bf 184}, 1 (1987).

\item{[26]} A. Kashlinsky, {\it Astrophys. J.} {\bf 331}, L1 (1988).

\item{[27]} E. V. Linder, {\it Astron. Astrophys.}, {\bf 206},
199 (1988).

\item{[28]} S. Cole and G. Efstathiou, {\it Mon. Not. Roy. Astron. Soc.}
{\bf 239}, 195 (1989).

\item{[29]} M. Sasaki, {\it Mon. Not. Roy. Astron. Soc.}
{\bf 240}, 415 (1989).

\item{[30]} K. Tomita and K. Watanabe, {\it Prog. Theor. Phys.}
{\bf 82}, 563 (1989).

\item{[31]} E. V. Linder, {\it Mon. Not. Roy. Astron. Soc.}
{\bf 243}, 353 (1990).

\item{[32]} L. Cay\'on, E. Mart\'inez-Gonz\'alez, and J. L. Sanz,
{\it Astrophys. J.} {\bf 403}, 10 (1993).

\item{[33]} L. Cay\'on, E. Mart\'inez-Gonz\'alez, and J. L. Sanz,
{\it Astron. Astrophys.} {\bf 284}, 719 (1994).

\item{[34]} B.A.C.C. Bassett, P.K.S. Dunsby, and G.F.R. Ellis, preprint
UCT-94-4 (1994); gr-qc/9405006.

\item{[35]} T. Fukushige, J. Makino, and T. Ebisuzaki, {\it Astrophys. J.}
{\bf 436}, L107 (1994); astro-ph/9409069.

\item{[36]} U. Seljak, preprint MIT-CSR-94-29 (1995); astro-ph/9505109.

\item{[37]} N. Kaiser, {\it Mon. Not. Roy. Astron. Soc.} {\bf 198}, 1033
(1982).

\item{[38]} L. Notalle, {\it Mon. Not. Roy. Astron. Soc.} {\bf 206}, 713
(1984).

\item{[39]} K.L. Thompson and E.T. Vishniac, {\it Astrophys. J.} {\bf 313},
517 (1987).

\item{[40]} M. Panek, {\it Astrophys. J.} {\bf 388}, 225 (1992).

\item{[41]} E. Mart\'inez-Gonz\'alez, J. L. Sanz, and J. Silk,
{\it Astrophys. J.} {\bf 355}, L5 (1990).

\item{[42]} M.J. Chodorowski, {\it Mon. Not. Roy. Astron. Soc.} {\bf 259},
218 (1992).

\item{[43]} E. Mart\'inez-Gonz\'alez, J. L. Sanz, and J. Silk,
{\it Phys. Rev.} {\bf D46}, 4193 (1992).

\item{[44]} E. Mart\'inez-Gonz\'alez, J. L. Sanz, and J. Silk,
{\it Astrophys. J.} {\bf 461}, 1 (1994); astro-ph/9406001.

\item{[45]} J.V. Arnau, M.J. Fullana, and D. S\'aez, {\it Mon. Not. Roy.
Astron. Soc.} {\bf 268}, L17 (1994).

\item{[46]} W. Hu and N. Sugiyama, {\it Phys. Rev. D} {\bf 50},
627 (1994); astro-ph/9310046.

\item{[47]} U. Seljak, preprint  MIT-CSR-95-13 (1995); astro-ph/9506048.

\item{[48]} R. Tuluie and P. Laguna, {\it Astrophys. J.} {\bf 445}, 73
(1995); astro-ph/9501059.

\item{[49]} L. Kofman and A. Starobinski, {\it Sov. Astron. Lett.} {\bf 11},
271 (1985).

\item{[50]} N. Gouda, N. Sugiyama, and M. Sasaki, {\it Prog. Theor. Phys.}
{\bf 85}, 1023 (1991).

\item{[51]} M. Kamionkowski and D.N. Spergel, {\it Astrophys. J.} {\bf 432},
7 (1994); astro-ph/9312017.

\item{[52]} N. Sugiyama and J. Silk, {\it Phys. Rev. Lett.} {\bf 73},
509 (1994); astro-ph/9406026.

\item{[53]} G. Jungman, M. Kamionkowski, A. Kosowsky, and D.N. Spergel,
preprint CU-TP-703 (1995); astro-ph/9507080.

\item{[54]} R. Tuluie, P. Laguna, and P. Anninos, preprint CGPG-95-96/10-1
(1995); astro-ph/9510019.

\item{[55]} E.F. Bunn and N. Sugiyama, {\it Astrophys. J.} {\bf 446}, 49
(1995); astro-ph/9407069.

\item{[56]} R.G. Crittenden and N. Turok, preprint PUPT 95-1569 (1995);
astro-ph/9510072.

\vfill\eject

\centerline{\bf Figure Caption}

\noindent{\it Figure One.}  This figure shows the observer at location
$x^\mu(\lambda_\O)$, the hypersurface of last scattering at
$\eta = \eta_\E$, and various paths connecting the two.  The true
geodesic in the perturbed metric is $x^\mu(\lambda)$, while the
background geodesic is $x^{\0\mu}(\lambda)$.  Adding the deviation
vectors $x^{\1\mu}(\lambda)$ and $x^{\2\mu}(\lambda)$ to the background
path yields increasingly accurate approximations to the true path.
The spacelike deviation vectors $p^{(a)i}$ are to be distinguished
from the $x^{(a)\mu}$, since the latter generally do not lie in
hypersurfaces of constant conformal time.

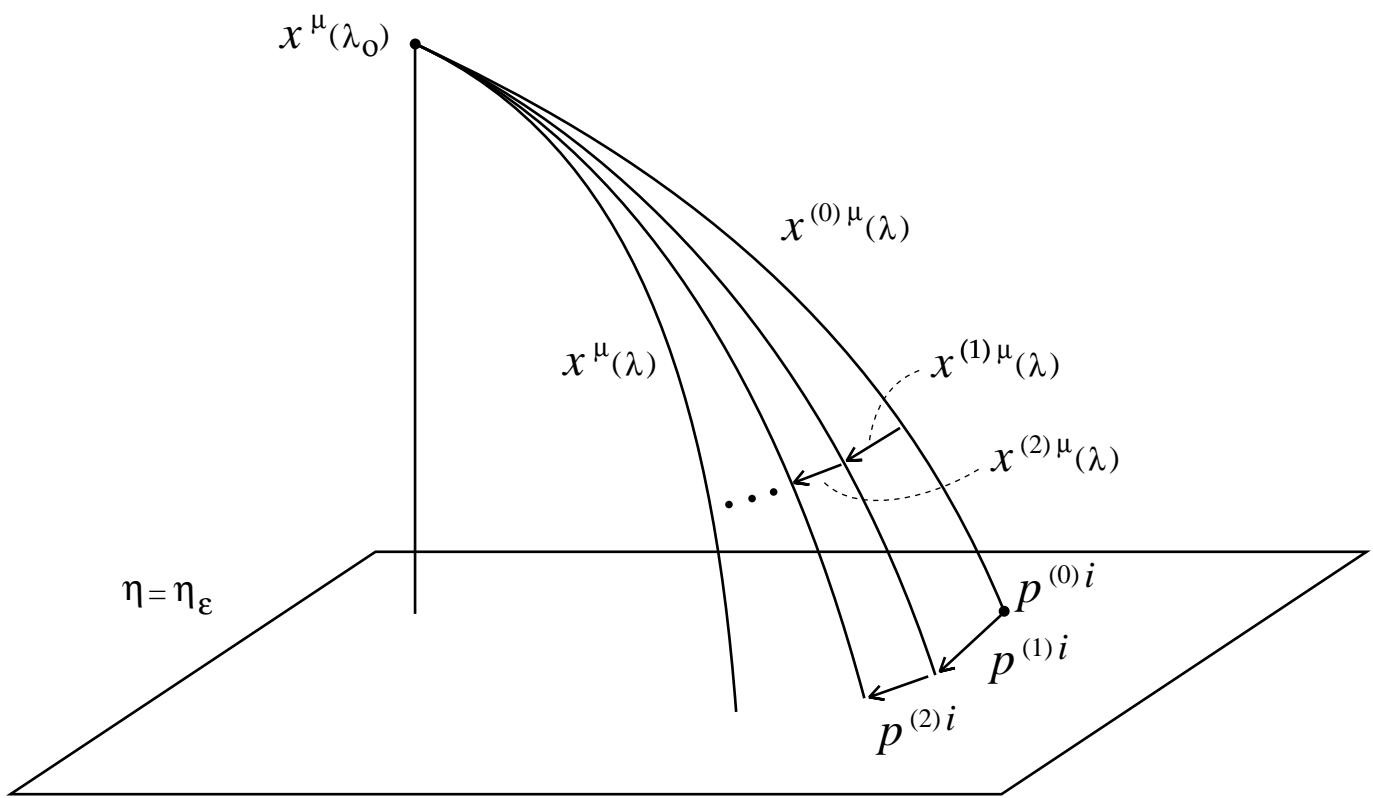
\bye